\documentclass[twocolumn]{aastex63}
\pdfoutput=1 
\usepackage{amsmath,amstext}
\usepackage{graphicx}
\usepackage[caption=false]{subfig}
\usepackage[figure,figure*]{hypcap}
\usepackage[hyphens]{url}
\usepackage{hyperref}
\received{June 1, 2020}
\revised{June 30, 2020}
\accepted{June 30, 2020}

\submitjournal{\apjs}

\shorttitle{Classifying Stellar Spectra}
\shortauthors{Roulston et al.}

\turnoffediting

\begin{document}

\title{Classifying Single Stars and Spectroscopic Binaries Using
Optical Stellar Templates}

\correspondingauthor{Benjamin Roulston}
\email{roulstbr@bu.edu}

\author[0000-0002-9453-7735]{Benjamin R. Roulston}
\altaffiliation{SAO Predoctoral Fellow}

\affiliation{Department of Astronomy, Boston University, 725 Commonwealth Avenue, Boston, MA 02215, USA}
\affiliation{Center for Astrophysics $\vert$ Harvard \& Smithsonian, 60 Garden St, Cambridge, MA 02138, USA}
\author[0000-0002-8179-9445]{Paul J. Green}
\affiliation{Center for Astrophysics $\vert$ Harvard \& Smithsonian, 60 Garden St, Cambridge, MA 02138, USA}
\author[0000-0002-3239-5989]{Aurora Y. Kesseli}
\affiliation{Leiden Observatory, Leiden University, P.O. Box 9513, 2300 RA, Leiden, The Netherlands}



\begin{abstract}
Stellar spectral classification is a fundamental tool of modern astronomy, providing insight into physical characteristics such as effective temperature, surface gravity, and metallicity. Accurate and fast spectral typing is an integral need for large all-sky spectroscopic surveys like the SDSS and LAMOST.  Here, we present the next version of PyHammer, stellar spectral classification software that uses optical spectral templates and spectral line index measurements. PyHammer v2.0 extends the classification power to include dwarf carbon (dC) stars, DA white dwarf (WD) stars, and also double-lined spectroscopic binaries (SB2). This release also includes a new empirical library of luminosity-normalized spectra that can be used to flux calibrate observed spectra, or to create synthetic SB2 spectra. We have generated physically reasonable SB2 combinations as templates, adding to PyHammer the ability to spectrally type SB2s. We test classification success rates on SB2 spectra, generated from the SDSS, across a wide range of spectral types and signal-to-noise ratios. Within the defined range of pairings described, more than $95\%$ of SB2s are correctly classified.
\end{abstract}



\section{Introduction}\label{sec:intro}

Spectroscopy is a cornerstone of modern astronomy. Our understanding of the composition of stars, molecular clouds, nebulae, and exoplanetary atmospheres are determined via spectroscopy. Spectroscopy is used to search for and characterize binary companions and exoplanets, to study the environments around stars and galaxies, and to map the distances to galaxies and quasars.

Stellar spectroscopy allows for the measurement of stellar radial velocities, key for the study of galactic archaeology and evolution, as well as for the study of binary stellar systems and their evolution. Modern spectroscopic instruments and telescopes can determine radial velocities from the tug of exoplanets around other stars. In the case of double-lined spectroscopic binaries (SB2), spectroscopy can reveal not only the presence of a binary system, but information on its individual components. Accurate and rapid stellar classification is crucial to extract physics from stellar spectroscopy, especially in the era of large surveys.


The current spectral classification system, the Harvard system, classifies stars into letter classes that follow a temperature scale. The current classes of O, B, A, F, G, K, M, L, T\added{\edit1{, Y}} represent stars and brown dwarfs across the stellar temperature range. O stars have the highest temperatures found in stars, and the \replaced{M, L, and T-dwarfs}{\edit1{M, L, T, and Y-dwarfs}} are the coolest (and most abundant, see \citealt{Bochanski2010}) spectral types which span the change from stars to brown dwarfs. While the majority of stars fall into these types, there are a few other common spectral types commonly found in large surveys, including the carbon (C) and white dwarf (WD) stars. The expanded Morgan-Keenan system \citep{MKK_sys} adds additional luminosity (i.e. giant, dwarf, sub-dwarf) classes to the spectral typing scheme. 

In recent years, there have been numerous implementations of automated spectral typing algorithms and software. These have included principal component analysis of large spectroscopic surveys like the Sloan Digital Sky Survey \citep[SDSS;][]{SDSS_4, McGurk2010}, neural networks \citep{Singh1998, Sharma2020},  fitting of synthetic spectra from model atmospheres, and comparisons to spectral line in stellar templates like PyHammer \citep{PyHammer_1}.

These automatic spectral typing methods have come about as a direct need of current and future large spectroscopic survey campaigns like SDSS and LAMOST \citep{LAMOST}. These surveys have already produced spectra for millions of stars, and with the beginning of the SDSS-V \citep{SDSS_5}, millions more will be observed with repeat, time-domain spectroscopy. These surveys represent an enormous, only partly-exploited, resource for astronomy. With the coming of age of time-domain astronomy and large scale, all-sky photometric surveys, like ZTF \citep{ZTF_1, ZTF_2, ZTF_3} and the Rubin Observatory Legacy Survey of Space and Time (LSST; \citealt{LSST}), the need for efficient and accurate stellar spectral typing is only going to become more relevant. 

In addition to large spectroscopic surveys, further advances are needed to spectrally type binary stars, particularly close interacting binary stars. Recent surveys have shown that almost half (46\%) of solar-type stars are in multiple systems \citep{Raghavan2010}. Higher multiplicity can be found for earlier-type stars, while later-type stars are found to have a lower multiplicity - near 27\% for M-dwarfs \citep{Winters2019}. Many of these systems are spatially unresolved and therefore undetected. However, the spectrum contains the light from both components and can tell us information about each one. In some cases, these components are of sufficiently different spectral type that the spectrum is visually striking as an SB2 (e.g. M+WD binaries). However, the majority of SB2 components are in spectral types that are closer together in the MK system (e.g. G+K). While there have been advances in `disentangling' SB2 spectra \citep{Spectangular_1, Spectangular_2}, these methods have generally relied on high resolution, high S/N, and multi-epoch spectra. These high quality spectra require significant dedicated telescope time.

Motivation for this work came from the combination of a large scale spectroscopic survey with a large fraction of stars being possible binary systems. The SDSS-IV's Time Domain Spectroscopic Survey (\citealp[TDSS;][]{Morganson2015, Ruan2016, MacLeod2018}, Anderson et al. 2020 in prep.) is a large spectral survey designed to collect optical spectra for a large sample of variable objects. The TDSS observed optical spectra for approximately 81,000 variable sources (quasars and stars) selected based on being spatially unresolved in SDSS imaging, and photometrically variable, without further regard e.g., for color or type of variability (Anderson et. al 2020, in prep.). One of the TDSS's main goals is the study of variable stars with a combination of spectroscopy and photometry. Approximately half of the periodic stellar systems in this sample are likely to be binaries, but do not show clear eclipses. These undetected binaries and their properties led to the motivation for the work detailed here. More details of the TDSS variable star survey can be found in Roulston et. al (2020, in prep.).

Here we present the newest version of PyHammer, a Python spectral typing suite. PyHammer has the advantage of needing only a single epoch of spectroscopy to perform spectral typing, including the new SB2 typing abilities detailed here. We also present a new library of empirically-derived, luminosity-normalized spectral templates. These luminosity spectra are used to create SB2 templates, which have also been added to PyHammer. This version has also been extended to include single carbon and DA white dwarf stars. 

\section{PyHammer} \label{pyhammer}

\subsection{PyHammer v1.0} \label{pyhammer1}

PyHammer \citep[][\url{https://github.com/BU-hammerTeam/PyHammer}]{PyHammer_1} is a Python based spectral typing suite that is based on the IDL program the Hammer \citep[][\url{http://myweb.facstaff.wwu.edu/~coveyk/thehammer.html}]{Hammer}. 

In v1.0 of PyHammer, spectral types were assigned by measuring spectral indices (similar to equivalent widths) for 34 atomic and molecular lines and comparing the measured indices to those of the templates. The best matching spectral type was selected as the one that minimized the $\chi^2$ difference between spectral indices. Because proper flux calibration across the full wavelength range of optical spectra can often be a significant challenge to achieve observationally, the use of spectral indices offers a distinct advantage for accurate classification of typical spectra.

The templates used by PyHammer are for single stars spanning types: O, B, A, F, G, K, M, L. Each of these classes contains a variety of sub-types and metallicities that are simultaneously compared. All of these templates were created from the co-addition of Sloan Digital Sky Survey \citep[SDSS;][]{SDSS_4} optical spectra, as detailed in \citet{PyHammer_1}.

The v1.0 PyHammer release extended the Hammer by including new templates to allow for spectral typing across different metallicities. It also provided a Python package that is easy to install and begin spectral typing without requiring IDL\footnote{\url{https://www.harrisgeospatial.com/Software-Technology/IDL}}.

\subsection{PyHammer v2.0: SB2} \label{subsec:phsb}

In our new release of PyHammer, available on GitHub\footnote{\url{https://github.com/BU-hammerTeam/PyHammer}}, we add two new single star spectral types for main sequence carbon stars \added{\edit1{(i.e. dwarf carbon, dC)}} and DA white dwarfs (WD), defining spectral indices for the C$_2$ and CH bandheads. The new dC stars span a range of broadband colors (and likely effective temperatures) from classic ``G'' to ``M'' type stars, while the WDs span a range of temperatures from 7000K to 100000K. 

While the Balmer line spectral indices were included in PyHammer v1.0, we include a second set that span wider Balmer line wavelength regions to help aid in the classification of the WDs. 

PyHammer v2.0 now also has the ability to detect some combinations of spectroscopic binaries. The details of the SB2 templates are discussed in Section \ref{sec:sb2}

\section{Carbon Star and White Dwarf Templates} \label{sec:single_templates}

\subsection{Carbon Star Templates}\label{subsec:c_single_temps}
\begin{singlespace}
\begin{deluxetable}{cDDDDD}
\tablecaption{dC Star Template Colors}
\tablewidth{1.0\textwidth}
\tablehead{\colhead{Template} & \twocolhead{$g - r$} & \twocolhead{$r - i$} & \twocolhead{$BP - RP$} & \twocolhead{N$_{spec}^\tablenotemark{a}$} & \twocolhead{S/N}}
\decimals
\startdata
dCG & 0.68 & 0.47 & 1.35 & 3 & 59 \\
dCK & 1.30 & 0.56 & 1.64 & 5 & 54 \\
dCM & 1.77 & 0.61 & 1.92 & 9 & 64 \\
\enddata
\tablenotetext{a}{\footnotesize Number of individual stellar spectra combined to create template.}
\tablecomments{Properties of the new dC star templates. Colors are the unweighted average of SDSS and Gaia colors of the component spectra and were used to help separate stars into the three dC star template classes (along with visual inspection). The templates for dCG, dCK, dCM correspond to `G', `K', and `M'-type dC stars respectively.} 
\end{deluxetable}
\label{tab:c_temp_colors}


\end{singlespace}

This release of PyHammer includes three new single star \added{\edit1{dwarf}} carbon (dC) star templates. These new dC star templates are made using dwarf carbon (dC) stars and were created in a similar method as the stellar library of \citet{PyHammer_1}, involving the co-addition of individual spectra to make each sub-type. 

The individual spectra used to make the dC star templates are from the SDSS sample of \citet{Green2013}. \citet{Green2013} identified carbon stars by visual inspection of single-epoch SDSS spectra compiled from the union of (1) SDSS DR7 spectra \citep{SDSS_DR7} having strong cross-correlation coefficients with the SDSS carbon star templates, and with (2) SDSS spectra with a DR8 pipeline class of STAR and subclass including the word carbon \citep{SDSS_DR8}. The subsample with main sequence luminosities (the dCs) were identified by their high proper motions.

The spectra used to make the new dC star templates were all selected from the SDSS DR16 \citep{SDSS_DR16}, which includes a combination of SDSS-I/SDSS-II and SDSS-III/SDSS-IV spectroscopic data. SDSS-I/SDSS-II spectra were taken with the legacy SDSS spectrograph, spanning a wavelength range of 3900 -- 9100\AA\ with a resolution of $R \sim 2000$. 
The newer eBOSS spectrograph \citep{SDSS_spectrograph} used in SDSS-III/SDSS-IV  covers the 3600 -- 10400\AA\ range at a resolution of $R \sim 2500$. 

From the \citet{Green2013} dC star sample, we made a series of quality cuts as follows: (1) SDSS $15.0 < r <17.0$ mag to ensure the SDSS sources are neither saturated nor have large uncertainties \citep{SDSS_filters} (2) average S/N $> 5$ for the SDSS spectrum (3) Gaia DR2 \citep{GaiaDR2} match within 2\arcsec\ (4) Gaia DR2 distance \citep{GaiaDR2_dist} S/N $> 5$\footnote{Our original cut of on parallax ($\varpi / \varpi_{\rm{error}} > 5$) did not translate to a distance S/N $> 5$ for all objects. The cut on the distance S/N ensures however that parallax and distance both have a S/N $> 5$.} (5) removed any dwarf Carbon (dC) stars known to be in binaries with a WD (i.e. those with evident DA/dC composite spectra \citealt{Heber1993, Liebert1994, Green2013, Si2014}) (6) removed any stars marked as giant in \citet{Green2013}.

After these cuts, we visually inspected each individual spectrum and removed those with bad flux regions and artifacts, which can happen due to background contamination, errors during the pipeline reduction, or a fiber not being correctly placed. During this visual inspection, the ``type'' of dC star was noted  (i.e., going progressively redder from `G' to `K' to `M'-types) based on continuum shape and strength of the CN \replaced{lines}{\edit1{bands}}.

We then placed the remaining dC stars into three groups based on the `type' given during the visual inspection. Then using the average SDSS colors of $g-r$, $r-i$ and Gaia $BP - RP$ we removed any sources that fell outside of the color locus for a given template. The breakdown of these colors can be found in Table \ref{tab:c_temp_colors}. The resulting templates (dubbed \replaced{C0, C1, C2}{\edit1{dCG, dCK, dCM}}) correspond approximately by color to  `G', `K', `M'-types, and were made from the co-addition of 3, 5, and 9 C star spectra respectively.

This co-addition follows the same method as used in \citet{PyHammer_1} for the original PyHammer; creating a wavelength grid that is logarithmically spaced (with 5 km s$^{-1}$ spacing), interpolating each component spectrum onto this grid, and then adding all the components together. The resulting template is then normalized so that the flux at 8000\AA\, is unity.

The new dC star templates are listed as [Fe/H] = 0.0, although their metallicity information is unknown. Both higher resolution spectra and well-tested model atmospheres would be needed, but are not yet available for dC stars.

\begin{figure*}
\centering
\plotone{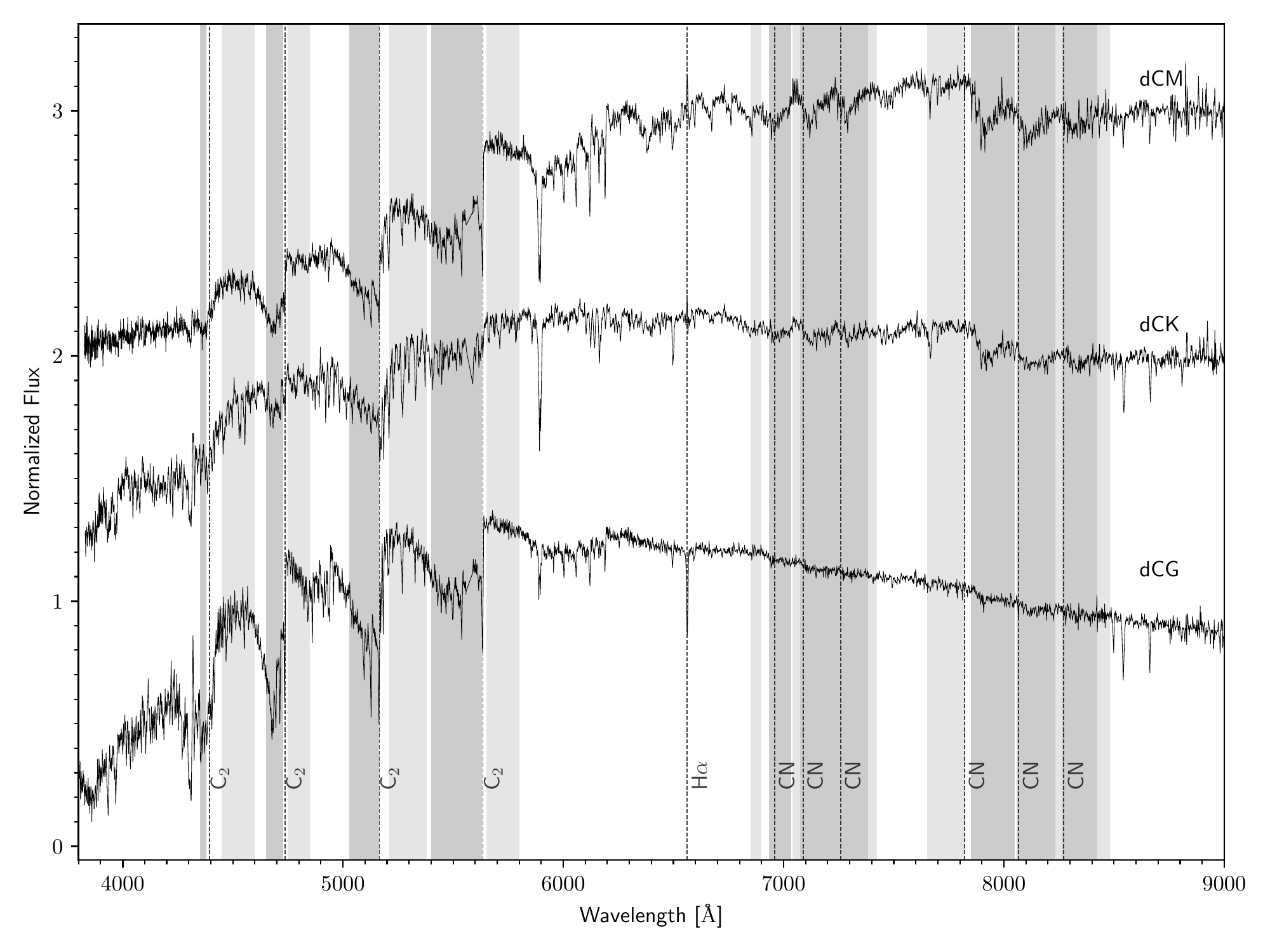}
\caption{New PyHammer single star dC star templates. Each of these sub-type templates is averaged from a sample of luminosity-normalized single epoch SDSS spectra. The striking and prominent C$_2$ and CN bandheads are visible across the dC stars. These bandheads, as well as H$\alpha$, are marked and labelled with dashed lines. We also show the new spectral index regions used by PyHammer for the automatic typing. There are 10 new dC star lines (4 C$_2$ and 6 CN), each consisting of a wavelength region within an absorption line or band, and a comparison 'continuum' region near the line region. The line regions and continuum regions are shown with dark and light gray shading, respectively. Each sub-type has been offset in flux for better visualization.}
\label{fig:ctemps}
\end{figure*}

The most striking features of dC stars are their prominent C$_2$ and CN bandheads. These can be seen in Figure \ref{fig:ctemps} which shows the three new dC star PyHammer templates. This figure also shows the variety of the dC class. The C$_2$ and CN molecular bandheads are marked, as well as the H$\alpha$ atomic line wavelength. These bandheads allow for accurate spectral typing, given that additional spectral indices are added to PyHammer. Details of these can be found in Section \ref{subsec:specInd}.

\subsection{WD Templates}\label{subsec:wd_single_temps}
\begin{singlespace}
\begin{deluxetable}{cDDDDDD}
\tablecaption{DA WD Templates}
\tablewidth{1.0\textwidth}
\tablehead{\colhead{Template} & \twocolhead{T$_{\rm{eff.}}$} & \twocolhead{$g - r$} & \twocolhead{$r - i$}  & \twocolhead{$BP - RP$} & \twocolhead{N$_{spec}^\tablenotemark{a}$} & \twocolhead{S/N}}
\decimals
\startdata
DA0.5 & 100 000 & -0.53 & -0.38 & -0.55 &   6  & 102 \\
DA1   &  50 000 & -0.53 & -0.37 & -0.55 &  12  & 120 \\
DA1.5 &  40 000 & -0.51 & -0.36 & -0.51 &  18  & 175 \\
DA2   &  30 000 & -0.47 & -0.34 & -0.45 &  61  & 301 \\
DA2.5 &  20 000 & -0.39 & -0.30 & -0.30 & 100  & 421 \\
DA3.5 &  15 000 & -0.30 & -0.25 & -0.15 &  99  & 325 \\
DA5   &  10 000 & -0.14 & -0.17 &  0.06 &  44  & 230 \\
DA5.5 &   9 000 & -0.02 & -0.07 &  0.24 &  28  & 184 \\
DA6.5 &   8 000 &  0.05 & -0.02 &  0.35 &  20  & 154 \\
DA7   &   7 000 &  0.19 &  0.05 &  0.52 &  16  & 164 \\
\enddata
\tablenotetext{a}{\footnotesize Number of individual stellar spectra combined to create template.}
\tablecomments{Properties of the 10 new DA WD templates. These templates span tempatures from 7000K to 100000K and all have a S/N above 100. The reported colors are from SDSS and Gaia DR2 where the value is the unweighted average of all the spectra used to make each template.} 
\end{deluxetable}
\label{tab:wd_temps}
\end{singlespace}

We have created new DA WD templates and added them to v2.0 of PyHammer. These WD templates were created using the same method as for the original single star PyHammer templates and new dC star templates. We used spectra from the \citet{Fusillo2019} WD sample, which used spectroscopically confirmed WDs from the SDSS to create selection criteria and color cuts to select WDs from Gaia DR2.

From this sample of 260000 high-confidence white dwarf candidates, we selected stars using the following quality cuts: (1) DA classification by \citet{Fusillo2019} (2) SDSS $15.0 <$ r $<17.0$\,mag (3) an existing SDSS spectrum with (4) mean S/N $> 5$ (5) a Gaia DR2 distance \citep{GaiaDR2_dist} S/N $> 5$. Similarly to the dC star templates, after these selection cuts were made each individual spectrum was visually inspected to check for bad flux regions and artifacts, removing stars with bad regions. The remaining stars were grouped by temperature, taken from model fits by \citet{Fusillo2019}, and co-added to create the 10 final DA templates, which were chosen to represent a reasonably-spaced temperature grid from 7000 to 100000\,K. Each individual spectrum was assigned to the template nearest in temperature (e.g. the T$_{\rm{eff.}} = 10000K$ template is made of WDs with $9500K < $ T$_{\rm{eff.}} \leq 12500$K).

\added{\edit1{The naming of the DA templates follow system and effective-temperature indicator introduced by \citet{Sion1983}, were we also follow the half-integer steps of \citet{Wesemael1993}.}}

Table \ref{tab:wd_temps} shows the resulting set of DA templates, their temperatures, and the number of individual spectra averaged to make them. As with the new dC star templates, these new DA templates are listed as [Fe/H] = 0.0. Although this is not valid for WDs, a metallicity value is required by the PyHammer software. 

\begin{figure*}
\centering
\plotone{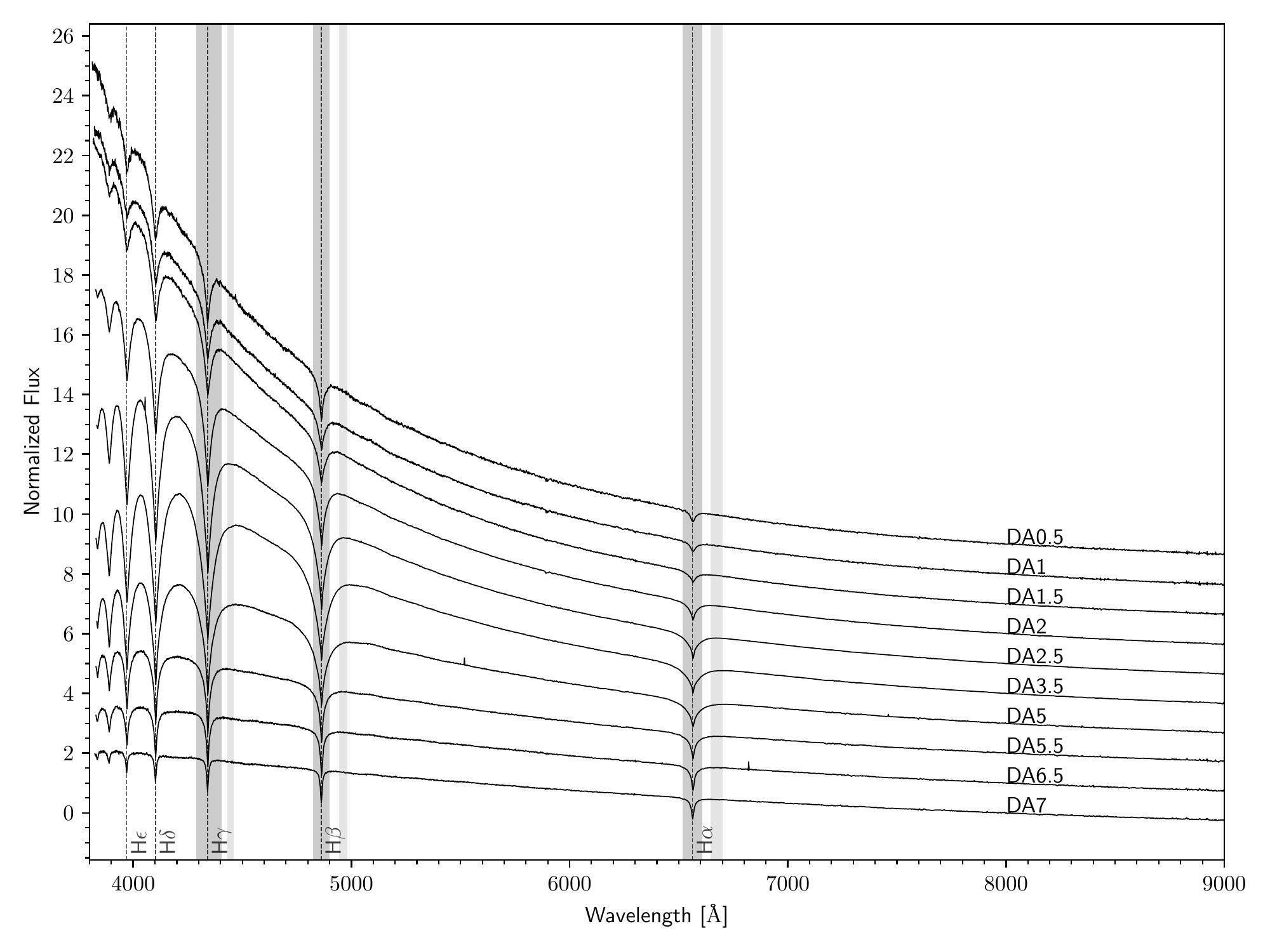}
\caption{New PyHammer DA WD star templates. Each of these sub-type templates is averaged from a sample of luminosity-normalized single epoch SDSS spectra. Each sub-type has been offset in flux for better visualization. The Balmer lines are marked and labelled with dashed lines. We also show the new spectral index regions used by PyHammer for the automated typing. Each spectral index consists of a line region and a continuum region near the line region. The line regions and continuum regions are shown with dark and light gray shading respectively.}
\label{fig:wdtemps}
\end{figure*}

Figure \ref{fig:wdtemps} shows these new DA single star PyHammer templates, illustrating the variety of the DA temperatures included.

PyHammer v1.0 used Balmer line indices for spectral typing; however, the featureless spectra of the new DA WDs were almost always confused with A and F star templates in v1.0. To distringuish WDs, additional Balmer line indices of varying widths were added as well as additional fitting methods. Details of these can be found in Section \ref{subsec:specInd}.

\subsection{PyHammer Spectral Indices}\label{subsec:specInd}

In addition to the new dC and DA WD templates, corresponding dC and DA line indices have been added to the list that PyHammer measures. The entire list, including new lines, is shown in Table \ref{tab:pyhammer2index}. This table shows the line and the comparison wavelength regions for the spectral index numerator and denominator. 

For the dC stars, we include the C$_2$ molecular bands in the blue and CN bands in the red. This allows for C bands to be calculated for either the bluer ``G-type'' carbon stars or the redder ``M-types''. 

We have added a second set of Balmer line indices specifically for the DA WDs. These new ``WD Balmer'' line indices add a wider wavelength region to the previously included narrow Balmer line indices. Since the DA WD Balmer lines are broadened due to strong pressure broadening, this helps both with line detection, and to distinguish the DA WD indices from those of main sequence stars with Balmer absorption.  However, these wider Balmer line indices alone were not enough to consistently differentiate between DA WDs and hot stars of A and F type. Therefore, an additional line width measurement for the H$\alpha$ line was added to the typing routine. This involved fitting the H$\alpha$ line with a Gaussian profile if the type is initially either A, F or DA WD. It then compares the fit width ($\sigma$) with that measured for most DA WDs; if the fit width is sufficiently large ($\sigma > 15$\AA), PyHammer classifies the spectrum as a DA.  
\begin{singlespace}
\begin{deluxetable*}{lcclcc}
\tablecaption{PyHammer v2.0 Spectral Indices}
\tablewidth{0.99\textwidth}
\tablehead{\colhead{Spectral Index} & \colhead{Numerator [\AA]} & \colhead{Denominator [\AA]} & \colhead{Spectral Index} & \colhead{Numerator [\AA]} & \colhead{Denominator [\AA]}}
\decimals
\startdata
\ion{Ca}{2} K              & 3924.8\,--\,3944.8 &  3944.8\,--\,3954.8 & CN $\lambda$7872           & 7850.0\,--\,8050.0 &  7650.0\,--\,7820.0 \\
Ca$\delta$                 & 4087.9\,--\,4117.9 &  4137.9\,--\,4177.2 & VO$\lambda$7912            & 7902.2\,--\,7982.2 &  8102.2\,--\,8152.2 \\
\ion{Ca}{1} $\lambda$4217  & 4217.9\,--\,4237.9 &  4237.9\,--\,4257.2 & CN $\lambda$8067           & 8059.0\,--\,8234.0 &  8234.0\,--\,8263.0 \\
G band                     & 4286.2\,--\,4316.2 &  4261.2\,--\,4286.2 & \ion{Na}{1}                & 8179.2\,--\,8203.2 &  8153.2\,--\,8177.2 \\
WD H$\gamma$               & 4290.0\,--\,4405.0 &  4430.0\,--\,4460.0 & CN $\lambda$8270           & 8263.0\,--\,8423.0 &  8423.0\,--\,8481.0 \\
H$\gamma$                  & 4333.7\,--\,4348.7 &  4356.2\,--\,4371.2 & TiO8                       & 8402.3\,--\,8417.3 &  8457.3\,--\,8472.3 \\
C$_2$ $\lambda$4382        & 4350.0\,--\,4380.0 &  4450.0\,--\,4600.0 & TiO $\lambda$8440          & 8442.3\,--\,8472.3 &  8402.3\,--\,8422.3 \\
\ion{Fe}{1} $\lambda$4383  & 4379.8\,--\,4389.8 &  4356.2\,--\,4371.2 & \ion{Ca}{2} $\lambda$8498  & 8485.3\,--\,8515.3 &  8515.3\,--\,8545.3 \\
\ion{Fe}{1}$\lambda$4404   & 4401.0\,--\,4411.0 &  4416.0\,--\,4426.0 & CrH-A                      & 8582.4\,--\,8602.4 &  8623.4\,--\,8643.4 \\
C$_2$ $\lambda$4737        & 4650.0\,--\,4730.0 &  4750.0\,--\,4850.0 & \ion{Ca}{2} $\lambda$8662  & 8652.4\,--\,8677.4 &  8627.4\,--\,8652.4 \\
WD H$\beta$                & 4823.0\,--\,4900.0 &  4945.0\,--\,4980.0 & \ion{Fe}{1} $\lambda$8689  & 8686.4\,--\,8696.4 &  8666.4\,--\,8676.4 \\
H$\beta $                  & 4848.4\,--\,4878.3 &  4818.3\,--\,4848.4 & FeH                        & 9880.0\,--\,10000.0 &  9820.0\,--\,9860.0 \\
C$_2$ $\lambda$5165        & 5028.0\,--\,5165.0 &  5210.0\,--\,5380.0 & VO$\lambda$7445            & 7352.0\,--\,7402.0, 0.5625  &  7422.0\,--\,7472.1    \\
\ion{Mg}{1}                & 5154.1\,--\,5194.1 &  5101.4\,--\,5151.4 & \mbox{ }                   & 7512.1\,--\,7562.0, 0.4375  &  7422.0\,--\,7472.1          \\
C$_2$ $\lambda$5636        & 5400.0\,--\,5630.0 &  5650.0\,--\,5800.0 & VO-B                       & 7862.2\,--\,7882.2, 0.5     &  7962.2\,--\,8002.2    \\
NaD                        & 5881.6\,--\,5906.6 &  5911.6\,--\,5936.6 & \mbox{ }                   & 8082.2\,--\,8102.2, 0.5     &  7962.2\,--\,8002.2          \\
\ion{Ca}{1} $\lambda$6162  & 6151.7\,--\,6176.7 &  6121.7\,--\,6146.7 & Rb-B                       & 7924.8\,--\,7934.8, 0.5     &  7944.8\,--\,7954.8    \\
WD H$\alpha$               & 6519.0\,--\,6609.0 &  6645.0\,--\,6700.0 & \mbox{ }                   & 7964.8\,--\,7974.8, 0.5     &  7944.8\,--\,7954.8          \\
H$\alpha$                  & 6549.8\,--\,6579.8 &  6584.8\,--\,6614.8 & Cs-A                       & 8498.4\,--\,8508.4, 0.5     &  8518.4\,--\,8528.4    \\
CaH2                       & 6815.9\,--\,6847.9 &  7043.9\,--\,7047.9 & \mbox{ }                   & 8538.4\,--\,8548.4, 0.5     &  8518.4\,--\,8528.4          \\
\cline{4-6}
CN $\lambda$6926           & 6935.0\,--\,7035.0 &  6850.0\,--\,6900.0 & Color region1 &  4160.0\,--\,4210.0 &  7480.0\,--\,7580.0 \\
CaH3                       & 6961.9\,--\,6991.9 &  7043.9\,--\,7047.9 & Color region2 &  4550.0\,--\,4650.0 &  7480.0\,--\,7580.0 \\
CN $\lambda$7088           & 7075.0\,--\,7233.0 &  7039.0\,--\,7075.0 & Color region3 &  5700.0\,--\,5800.0 &  7480.0\,--\,7580.0 \\
TiO5                       & 7127.9\,--\,7136.9 &  7043.9\,--\,7047.9 & Color region4 &  9100.0\,--\,9200.0 &  7480.0\,--\,7580.0 \\
CN $\lambda$7259           & 7233.0\,--\,7382.0 &  7382.0\,--\,7425.0 & Color region5 & 10100.0\,--\,10200.0 &  7480.0\,--\,7580.0 \\
VO$\lambda$7434            & 7432.0\,--\,7472.0 &  7552.0\,--\,7572.0 & \mbox{ } &\mbox{ } &  \mbox{ } \\
\enddata
\tablecomments{Spectral indices for v2.0 of PyHammer. For the indices that have two numerator regions the weight for each numerator is shown. Atomic and molecular lines are in order of increasing wavelength. Some  Color region indices, separated from the atomic and molecular line indices at the bottom of the right set of indices, are also in increasing wavelength order.} 
\end{deluxetable*}
\label{tab:pyhammer2index}

\end{singlespace}

\section{Luminosity Stellar Templates} \label{sec:lumspec}

To create single star templates that can then be combined to create realistic spectroscopic binary templates, we built a library of luminosity-normalized spectra\added{\edit1{\footnote{Here luminosity-normalized refers to spectra in luminosity units, not to be confused with luminosity classes from the MK system.}}}, in units of erg s$^{-1}$ \AA$^{-1}$. To do this, we sought optical spectral libraries with precise flux calibrations that allow transformation into luminosity units using well-measured distances.

We created this by selecting O, B, A, and F \added{\edit1{main sequence}} stars from \citet{Pickles1998}, G, K, and M stars from the SDSS-IV MaStar program \citep{MaStar} program, dC stars from \citet{Green2013}, and DA WDs from \citet{Fusillo2019}.

The MaStar survey uses fiber bundles, which can achieve much more accurate flux calibration than the normal SDSS survey. However, the MaStar sample (DR 16) lacks  O, B, A, and F stars that meet our quality cuts. For those spectral types, we used the \citet{Pickles1998} library. This library also is well flux-calibrated and has a similar resolution to SDSS and MaStar (R $\sim 2000$).
  
There are no public digital libraries of precisely flux-calibrated C and WD star spectra. For these spectral types, we used the same libraries that we made their single star templates from. Our Gaia distance quality cuts ensure accurate distances to perform the flux to luminosity transformation. 

The O, B, A, and F stars from \citet{Pickles1998} have excellent relative flux calibrations, but are presented in normalized units where each spectrum is normalized to 1.0 at 5556\AA.  Since absolute magnitudes M$_V$ are reported for each, we used the $V$ band filter response function from \cite{Bessell1990} to perform synthetic photometry, thereby finding the appropriate scale factor to convert these templates into luminosity units of erg s$^{-1}$ \AA$^{-1}$.

For the G, K, and M spectral types we matched each MaStar star to Gaia DR2 and selected the best spectrum for each spectral type and sub-type. This best spectrum was chosen as having the best combination of Gaia DR2 parallax S/N and Gaia $G$ magnitude S/N. After selecting those with Gaia DR2 $\varpi / \varpi_{\rm{err}} > 10$, we then chose the best S/N spectrum in each sub-type bin. Then, using the Gaia DR2  distance, we converted these flux spectra into luminosity spectra in units of erg s$^{-1}$ \AA$^{-1}$.  

For the dC and DA WD stars, we used a similar method to the GKM stars. However, since these objects are from the main SDSS-IV survey, some may have poorer absolute flux calibrations due e.g., to sub-optimal individual fiber placement or transmissivity. We mitigate this by averaging.  We converted each of the individual spectra for each template into luminosity units using the Gaia DR2 distances. Then, we co-added and averaged to get an average luminosity spectrum for each spectral type.  

Although PyHammer contains single L templates, we do not make L star spectral luminosity templates, because the L templates are constructed from very faint spectra ($r > 21$), outside our range of quality criteria.  These L spectra likely have poor flux calibrations that are not suitable for transforming into luminosity units. 

This luminosity-normalized digital spectral library allows for a variety of useful applications. The templates can be combined to create templates for spectroscopic binaries as described in the next section. Another important application is using these templates for flux calibration. Once an observed spectrum has been typed using PyHammer, one can divide the appropriate template by the observed spectrum, fit with a low-order polynomial, and then multiply the polynomial by the observed spectrum to get a luminosity-normalized observed spectrum.

\begin{figure*}
\centering
\epsscale{1.2}
\plotone{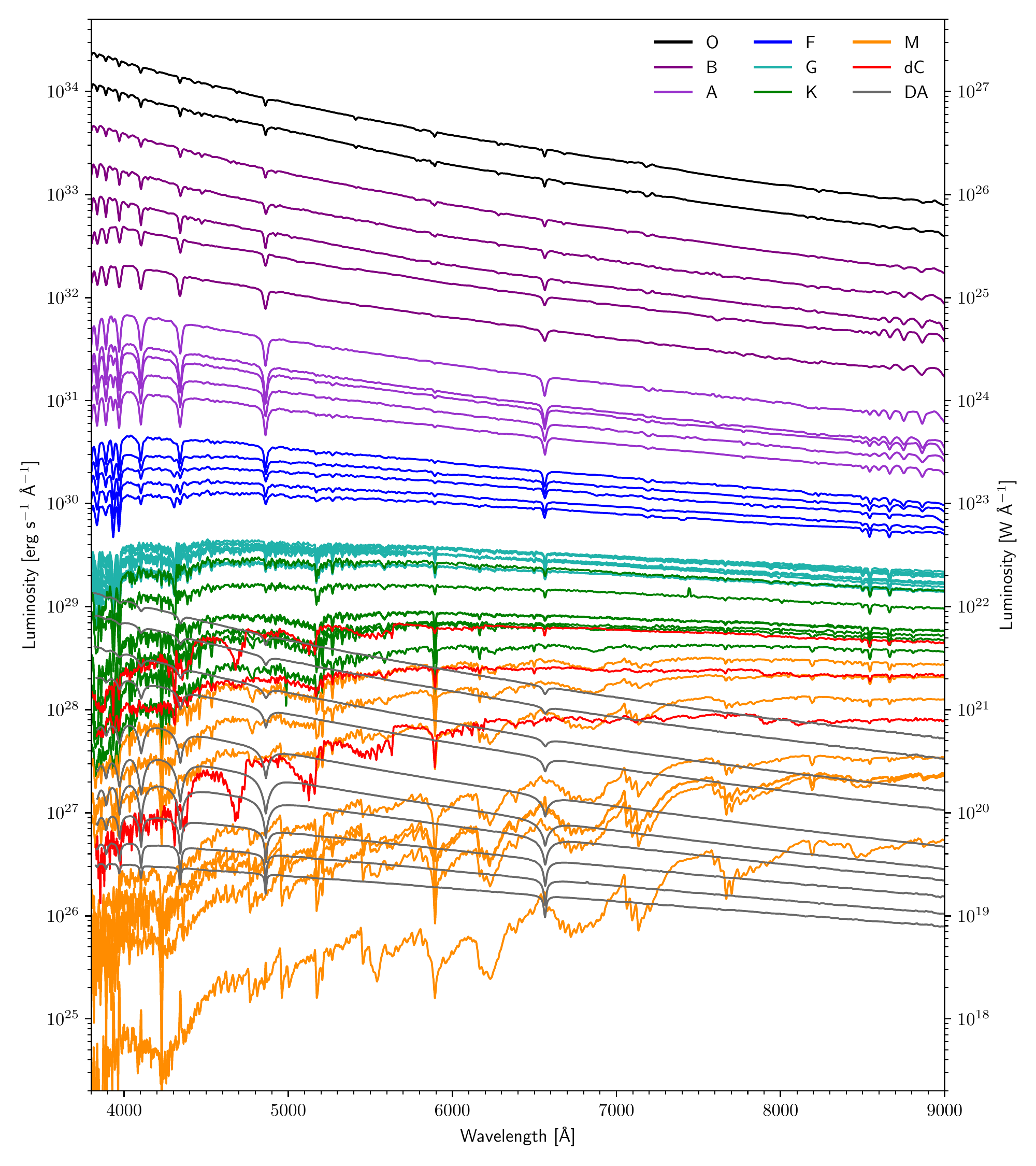}
\caption{New PyHammer luminosity-normalized stellar templates. These templates are made using digital stellar spectra from  \citet{Pickles1998} library, the MaStar library and dC and DA WD stars from \cite{Green2013} and \citet{Fusillo2019} respectively. The MaStar, dC, and WD libraries all use flux-calibrated spectra and are converted to luminosity units using Gaia DR2 distances. The \citet{Pickles1998} library is converted to luminosity units using reported M$_V$ and synthetic photometry. Note that the luminosity axis is spaced logarithmically to clearly show the variation of stellar luminosty with spectral type.}
\label{fig:lumspec}
\end{figure*}

Figure \ref{fig:lumspec} shows all of the luminosity spectra from our library that we then used to create spectroscopic binary templates. All of the luminosity spectra are in units of erg s$^{-1}$ \AA$^{-1}$ and have been smoothed by a boxcar of 10 pixels to aid in visualization.

This luminosity normalized spectral library can be found on Zenodo\footnote{\replaced{\url{https://doi.org/10.5281/zenodo.3900328}}{\edit1{\dataset[https://doi.org/10.5281/zenodo.3871959]{\doi{10.5281/zenodo.3900328}}}}} in FITS format.

\section{Spectroscopic Binary Templates}\label{sec:sb2}

Using the luminosity library from Section \ref{sec:lumspec}, we were able to combine these spectra to create a library of \added{\edit1{main sequence}} spectroscopic binary templates. This can be done by adding the component spectra together on a common wavelength axis to form a combined SB2 spectrum (the common wavelength axis we use is the PyHammer template wavelength grid which is logarithmically spaced between 3550\AA\ and 10500\AA\ with spacing of 5 km s$^{-1}$).

Not all combinations \added{\edit1{of our main sequence luminsioty templates}} make useful SB2 templates, as the more luminous stars easily overpower the faintest ones (e.g., an A5+M2 binary would be useless as the A star would be almost 10$^3$ times more luminous than the M star and no M star features would be visible).

To limit the combinations to those with some realistic hope of detection, we require that at least 20\% of the pixels of the two constituent spectra be within 20\% of the luminosity of each other.

For practical reasons of classification accuracy detailed in
Section \ref{sec:accuracy}, we only build SB2 combinations from constituents of different main spectral types (i.e. no A+A, F+F etc.). This results in the following combinations of main SB2 spectral types: A+F, F+G, F+K, G+K, G+dC, G+DA, K+M, K+dC, K+DA, M+dC, M+DA, dC+DA.

We have created and included these new SB2 templates to allow PyHammer the ability to spectral type SB2s based on a single epoch of optical spectroscopy. The SB2s generated and studied in the current work do not include any giant stars.

\subsection{SB2 Radial Velocities} \label{SB2rvs}

In addition to spectral typing, PyHammer has the ability to measure the radial velocity (RV) of an input spectrum. As detailed in \citet{PyHammer_1}, PyHammer uses a cross-correlation method across three wavelength regions. \citet{PyHammer_1} report that the original PyHammer has a RV accuracy of 7--10 km s$^{-1}$ for mid-temperature and low-temperature stars and 10--15 km s$^{-1}$ for high-temperature stars.

In the process of this work, we also considered adding the ability to PyHammer to fit the radial velocity (RV) of each of the SB2 component spectra. This would involve using our luminosity spectral library to create SB2 composite templates on the fly, fitting the SB2 to the input spectrum with the RVs for both components as free parameters.  This would be useful to find SB2s with components of similar spectral type (e.g. M2+M3 or F5+F6 etc.) where PyHammer v2.0 likely will classify the system as a single star. However, in such cases, there may often be widening or even separation of spectral lines due to the radial components of orbital motion of the components, potentially allowing RV fitting to detect the RVs of both stars. 

In practice, however, this proved difficult for a variety of reasons.  Mainly, the S/N of most SDSS spectra is not high enough to allow for this robust of a fitting routine, and our attempts at recovering simulated RV shifts were unsuccessful.
After implementing and testing a few methods, our retrieved RV measurements for simulated SB2s have, so far, been unreliable and so we do not include this tool in PyHammer v2.0. Stars determined to be best fit by a SB2 template will have a RV reported by the software as NaN. The ability to calculate RVs for single stars however remains the same as with the original PyHammer (including for the new C and WD templates).

\section{Accuracy} \label{sec:accuracy}

Our initial SB2 templates included all templates those met the requirement that 20\% of the pixels of the two constituent spectra be within 20\% of the luminosity of each other, including SB2s wherein both the primary and secondary were of the same main spectral type (e.g. A2+A3, M2+M4). 

However, after initial accuracy tests, we found that the classification accuracy rates for single stars dropped significantly when SB2s with the same main spectral type were included in PyHammer. For example, a single F5 star is unlikely to be misclassified as an F2+G2 SB2, but is quite likely to be misclassified as e.g., an F2+F5 SB2.

\begin{figure}
\centering
\epsscale{1.2}
\plotone{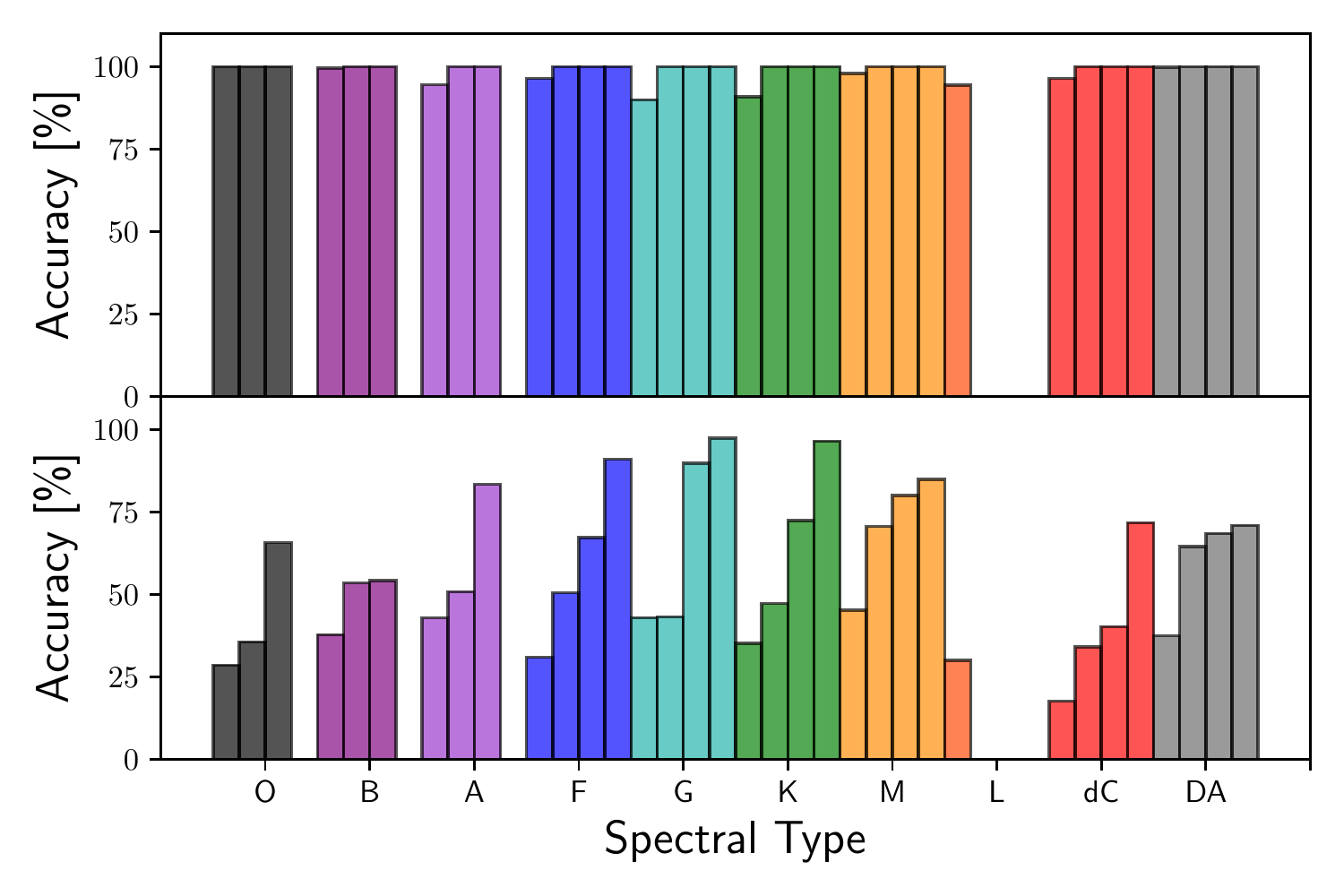}
\caption{Classification accuracy for single stars for two different sets of SB2 combinations. The lower panel shows the accuracy rates for single stars typed as single stars when including same main spectral type SB2s (e.g. A+A, or G+G). The upper panel shows the accuracy rates for single stars typed as single stars when same main spectral type binaries are excluded from PyHammer. For each spectral type there are 4 bars for 4 S/N bins (left to right): S/N $< 5$, $5 \leq$ S/N $ < 10$, $10 \leq$ S/N $ < 20$, S/N $\geq 20$. Note that not all spectral types have all S/N bins. Given the strong degradation in accuracy for single stars when allowing same main type SB2s, we exclude them from the software.}
\label{fig:singel_acc_vs_allcombo}
\end{figure}

Figure \ref{fig:singel_acc_vs_allcombo} shows the classification accuracy for single stars being typed as single stars when including SB2s with the same main spectral type. The bottom panel shows the accuracy when including same main spectral type SB2s and the top panel shows the accuracy rate when excluding same main spectral type SB2s. This figure shows how strongly the single star accuracy rates are affected by including same main spectral type SB2s. 

One possible reason that these same main spectral type SB2s negatively affect the single star accuracy rates is that there are two nearly equivalent templates in terms of spectral indices. For example, when typing an F5 spectrum it could be equally well-matched to an F5 template or an F4+F5 template. This results in single stars that have lower S/N or a noisy spectrum to be best typed by a SB2 with the same main spectral types. 

For this reason, we do not include these same main spectral type SB2s templates in PyHammer v2.0, limiting SB2s to have different main spectral types. Classification accuracies discussed further in this section refer to the accuracy of PyHammer using only the SB2s which have different main spectral types. 

However, all possible SB2 combinations meeting the 20\% criteria outlined in Section \ref{sec:sb2} are included in the Zenodo library for completeness, whether or not they combine the same main spectral types. This includes some types, like dC+dC, which would be expected to be extremely rare in the cosmos for reasons of stellar evolution. As well as DA+DA types, which could be extremely interesting but difficult to detect with PyHammer.

We have tested our new SB2 templates for their accuracy\footnote{The accuracies detailed here are representative of SDSS spectra and may not reflect results for spectra of different resolution, wavelength coverage, relative flux calibration, or quality.} and their dependence on the input spectral S/N.  We tested accuracies for all templates across a range of S/N.  We did this by degrading each template by varying levels of noise. We created a Gaussian distribution for each pixel centered at the pixel's flux with the standard deviation given by an integer multiple of the template error at that pixel. We then used these distributions to draw new noisy spectra for integer multiples between 1$\sigma$ and 50$\sigma$. Using PyHammer, we typed each noisy test spectrum at each degradation level. To better represent the accuracies for different use cases, we selected three ``criteria'' of classification accuracy to test. 

The first criterion (criterion 0) is the least stringent, allowing any combination of sub-types as long as the two main spectral types are correct.\footnote{Accuracy criterion 0 may be imprecise because it has discontinuous jumps at spectral type boundaries. For example, a F9+G9 classified as F9+K0 would be incorrect, even though a G9 is just one sub-type away from K0. However, this affects only a small percentage of our SB2 combinations.} For example, a M2+DA3.5 would be correct even if labeled as a M1+DA7 because the main spectral types are correctly M and DA, but if it were labelled as K7+DA0.5, it would be counted as incorrect. 

The second criterion (criterion 1) increases the requirements to count correct typing as only those SB2s classified by our code to be within one sub-type, in either or both of the components. In this case, a M2+DA3.5 would be counted correct if labelled as M2+DA2.5 (or M3+DA5, etc.), but would be incorrect if labelled as \replaced{M3+WD6}{\edit2{M2+DA6}}.

The third and most stringent criterion (criterion 2) counts the classification as correct if (and only if) the exact spectral types and sub-types for both components of the SB2 are correct. An example is a M2+DA3.5 would be classified correct only if labelled as M2+DA3.5; if labelled as a \replaced{M2+WD2.5}{\edit2{M2+DA2.5}}, it would be incorrect.

\begin{figure*}
\centering
\epsscale{0.9}
\plotone{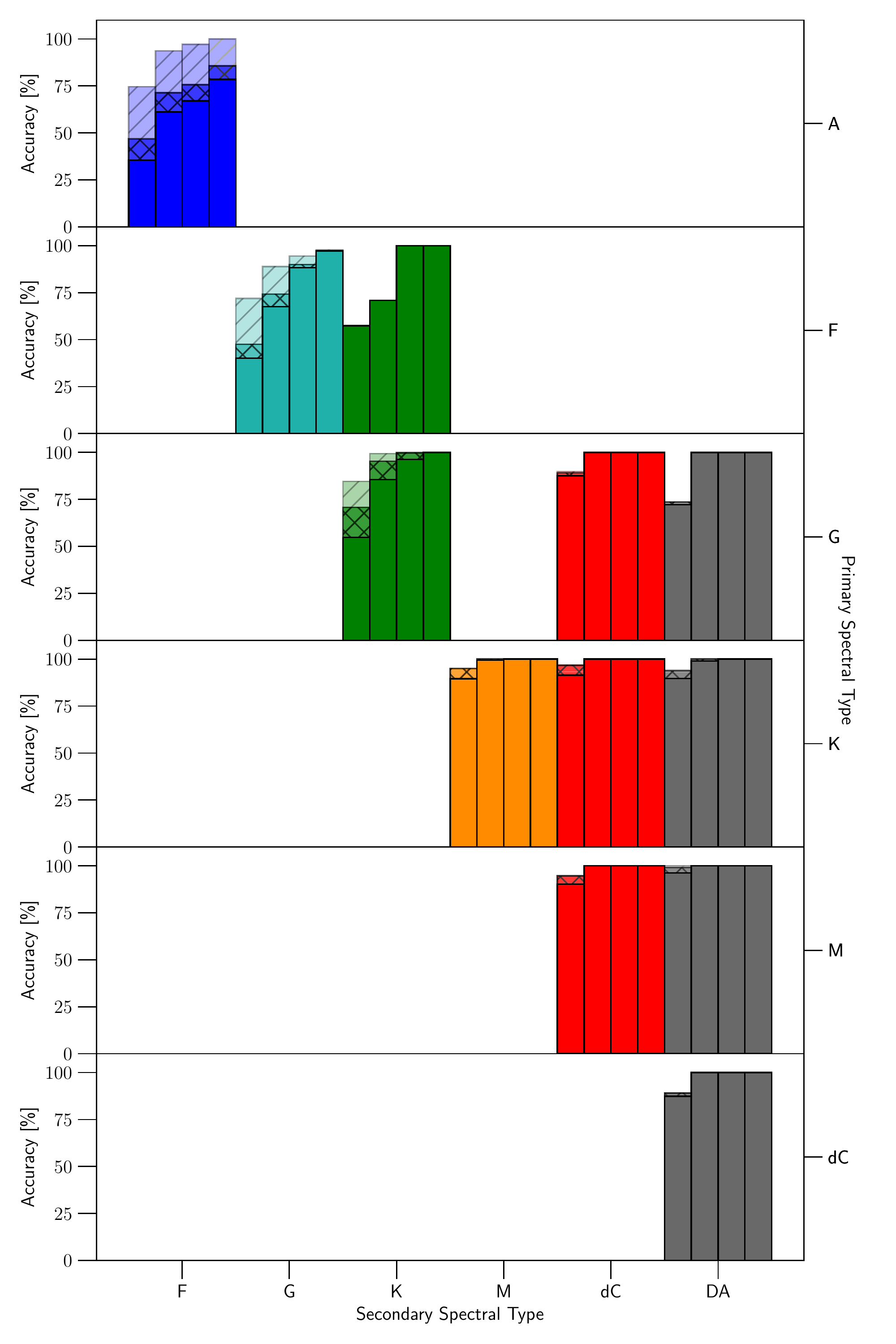}
\caption{SB2 accuracy based on sub-type, S/N, and accuracy criteria. Each primary+secondary combination has 4 bars for 4 S/N bins (left to right): S/N $< 5$, $5 \leq$ S/N $ < 10$, $10 \leq$ S/N $ < 20$, S/N $\geq 20$. Each bar then has 3 stacked components representing the previously described accuracy criteria (using criterion 0, 1, or 2). criterion 0 is represented by the most transparent bars (single diagonal hatching), criterion 1 by the middle transparent bars (double diagonal hatching), and criterion 2 by solid color bars (no hatching).}
\label{fig:sb2_acc}
\end{figure*}

Figure \ref{fig:sb2_acc} shows the accuracy for each of the SB2 groups in PyHammer 2.0. Each of the panels shows the accuracy for one of the 6 possible primary spectral types (A, F, G, K, M, dC). Each panel then shows the accuracies for the possible combinations of secondary types (e.g. A+F, or dC+DA).  Each primary+secondary combination has 4 bars for 4 S/N bins (left to right): S/N $< 5$, $5 \leq$ S/N $ < 10$, $10 \leq$ S/N $ < 20$, S/N $> 20$. Each bar then has 3 stacked components representing the previously described accuracy criteria (0, 1, or 2). Criterion 0 is represented by the most transparent (single diagonal hatching), criterion 1 by the partially transparent (double diagonal hatching), and criterion 2  by the solid color (no hatching) bars.

From Figure \ref{fig:sb2_acc}, we see that PyHammer's accuracy with SB2 stars is dependent both on the input spectrum's S/N and on the spectral type combination. As expected, the lowest S/N has the lowest accuracy, which holds across all three ``criteria'' for counting accuracy. We also can see that early type stellar combinations (i.e. A+F, F+G, F+K)  tend to be less reliable. This is expected, as the early types of A and F are spectrally similar, with the main features being the Balmer lines. In contrast, late type combinations (e.g. G+dC, K+M, M+dC, etc.) are much more - in some cases nearly 100\%  - accurate. This is likely due to the strong difference in visible atomic and molecular lines and bands within these spectral types (e.g. TiO bands in M dwarfs, and CN, CH, and C$_2$ bands in dC stars). PyHammer is particularity good at identifying binaries of late type stars with a DA companion (i.e. G+DA, K+DA, M+DA, C+DA) due to the strong WD Balmer lines in the blue with strong late type stellar features in the red. These types are all nearly 100\% accurate across all S/N bins and accuracy criteria.   

\begin{figure*}
\centering
\epsscale{1.2}
\plotone{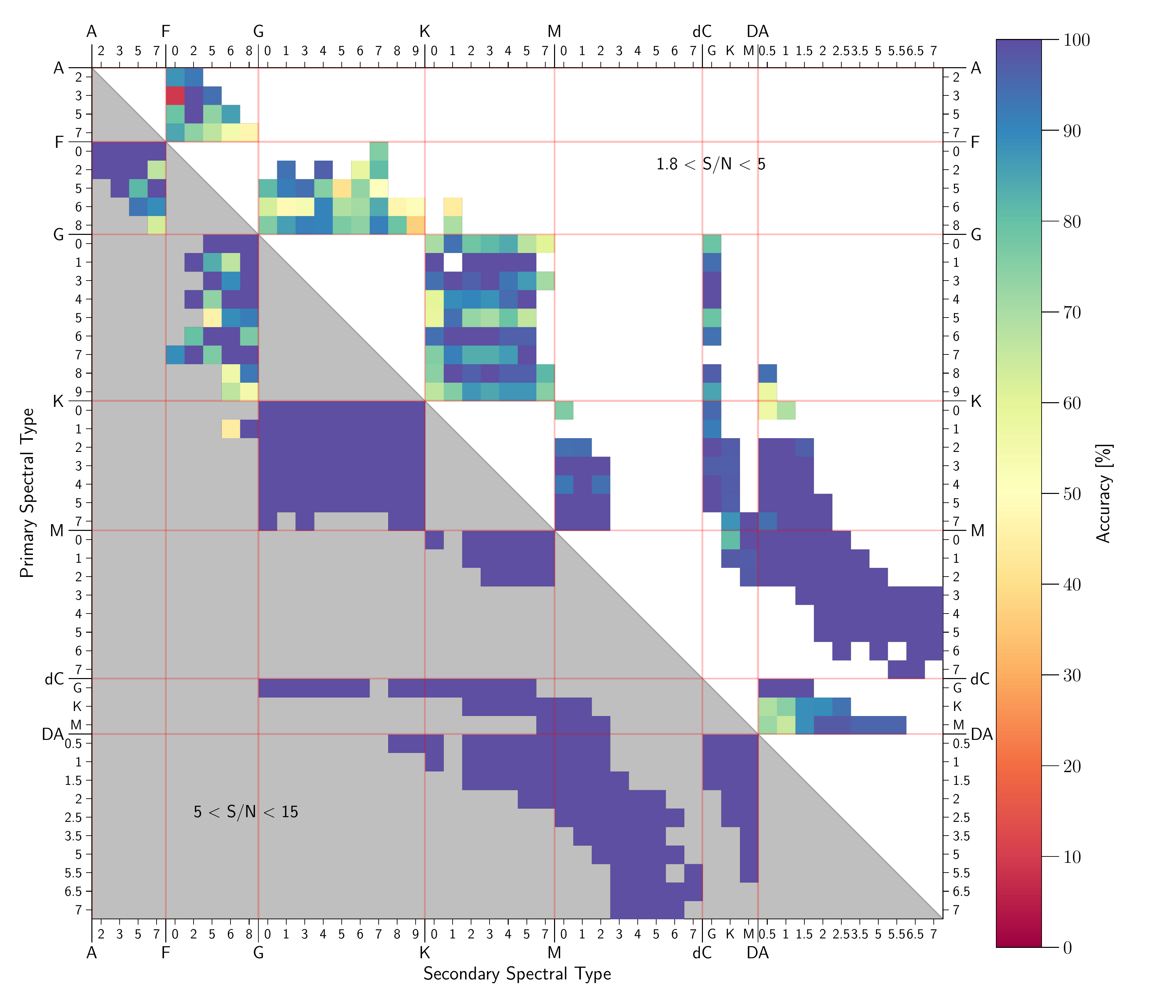}
\caption{SB2 accuracy based on accuracy criterion 0 (i.e. only main-types need be correct). The upper triangle (white background) of the plot shows the accuracy rates for spectra with $1.8 < $ S/N $< 5$, and the lower triangle (gray background) plot shows accuracy rates for spectra with $5 < $ S/N $< 15$. All possible SB2 combinations are shown in this figure, with the primary type and sub-type labelled along the y-axis and the secondary type and sub-type along the x-axis. The percentage accuracy for a specific SB2 combination is derived from all the degraded test spectra for that SB2 type that fall within the given S/N bin. }
\label{fig:conf_matrix}
\end{figure*}

Figure \ref{fig:conf_matrix} shows the accuracy for specific SB2 combinations. These accuracies are for criterion 0 (main types correct) and are an average of the degraded test spectra for that SB2 type that fall within the given S/N range. The figure shows two S/N ranges, with the $1.8 < $ S/N $< 5$ bin given above the diagonal in the upper triangle, and the $5 < $ S/N $< 15$ bin given below the diagonal in the lower triangle. This figure shows again that late-type combinations and those combinations with a DA WD component tend to be the most accurate at low S/N. However, at higher S/N (above $\sim$10) most combinations are above 90\% accurate in all three criteria of accuracy.

We also report the accuracy of PyHammer v2.0 in identifying between the single star and SB2 star templates. Table \ref{tab:accuracy} shows these accuracy rates between the single star and SB2 star classes. These rates are calculated from the total average across all S/N and across all spectral types and SB2 combinations. Here, an accurate typing is counted whenever a single star is typed as a single star or an SB2 is typed as a SB2. All other combinations are counted as incorrect (i.e., a single star classified as an SB2 or an SB2 classified as a single star). 

These accuracies give the rates at which, on average, we expect PyHammer to mistype between single and SB2 star templates. There is a dependence on both spectral type and S/N, with these misclassifications all occurring for S/N $< 5$ and 32\% being for A or F types. Misclassifications of A and F types are again not surprising, as those classes are very similar with predominant Balmer lines only. With low S/N, it is hard for PyHammer to distinguish between a low S/N A star and an A+F SB2. Overall, as shown in Table \ref{tab:accuracy}, PyHammer has about 95\% accuracy in correctly identifying between the single and SB2 star classes. 

\begin{singlespace}
\begin{deluxetable}{lDD}
\tablecaption{In Class Accuracy Rates}
\tablewidth{1.0\textwidth}
\tablehead{\colhead{SB2 Type} & \twocolhead{N$_{spec}$}       & \twocolhead{Accuracy}}
\decimals
\startdata
A+F  &  424 &  90.3 \\
F+G  &  769 &  90.8 \\
F+K  &   55 &  94.6 \\
G+K  & 1665 &  88.2 \\
G+dC  &  253 &  94.5 \\
G+DA &   72 &  88.9 \\
K+M  &  375 &  95.7 \\
K+dC  &  433 &  98.4 \\
K+DA &  580 &  95.7 \\
M+dC  &  203 &  95.1 \\
M+DA & 1686 & 100.0 \\
dC+DA &  708 &  90.4 \\
\hline
SB2 Average & 602 & 93.5 \\
\hline
\hline
Single Type & . & . \\
\hline
O  &  187 & 100.0 \\
B  &  421 &  99.5 \\
A  & 1406 &  94.5 \\
F  & 2140 &  96.4 \\
G  & 2278 &  90.1 \\
K  & 1653 &  91.0 \\
M  & 1507 &  97.9 \\
L  &  250 &  94.4 \\
dC  &  137 &  96.4 \\
DA &  412 &  99.8 \\
\hline
Single Average & 1039 & 96.0 \\
\enddata
\tablecomments{Accuracy and misclassification rates between the single and SB2 star classes. These rates are for S/N $< 5$ and only represent the accuracy and error rates between the single star and SB2 star template groups (i.e., this does not represent the total true accuracy rate because it only accounts for errors between the single and SB2 star classes). For each spectral type shown, the rate shown is the percentage of all test spectra in that primary-secondary type bin that were classified correctly as a SB2 for SB2 stars (and single for single stars).} 
\end{deluxetable}
\label{tab:accuracy}
\end{singlespace}

\section{PyHammer GUI} \label{gui}

The graphical user interface (GUI) for PyHammer SB2 is functionally similar to the GUI in v1.0 of PyHammer. We have made a few minor updates and included the new functionality needed for classification using the SB2 templates. 

PyHammer uses a $\chi^2$ method to compare the spectral indices of the input spectrum to that of the templates. We now show and report this raw ``distance'' measure on the GUI screen as LineDist to aid users when visually checking the classifications. Along with this statistic, we also report the residual between the chosen template and the input spectrum as well as the residual weighted by the errors. These allow the user to easily see the statistical change in the fit of each template in addition to a visual check.

There are now two additional sliders and a toggle for SB2 templates. The toggle will turn on to include both single and SB2 templates, or off to include only single star templates. The additional sliders allow the user to select a secondary star's spectral type and sub-type based on the selected primary types. Only valid combinations from our list of SB2 templates are allowed, with unavailable options greyed out.

\begin{figure*}
\centering
\includegraphics[angle=90,scale=0.37]{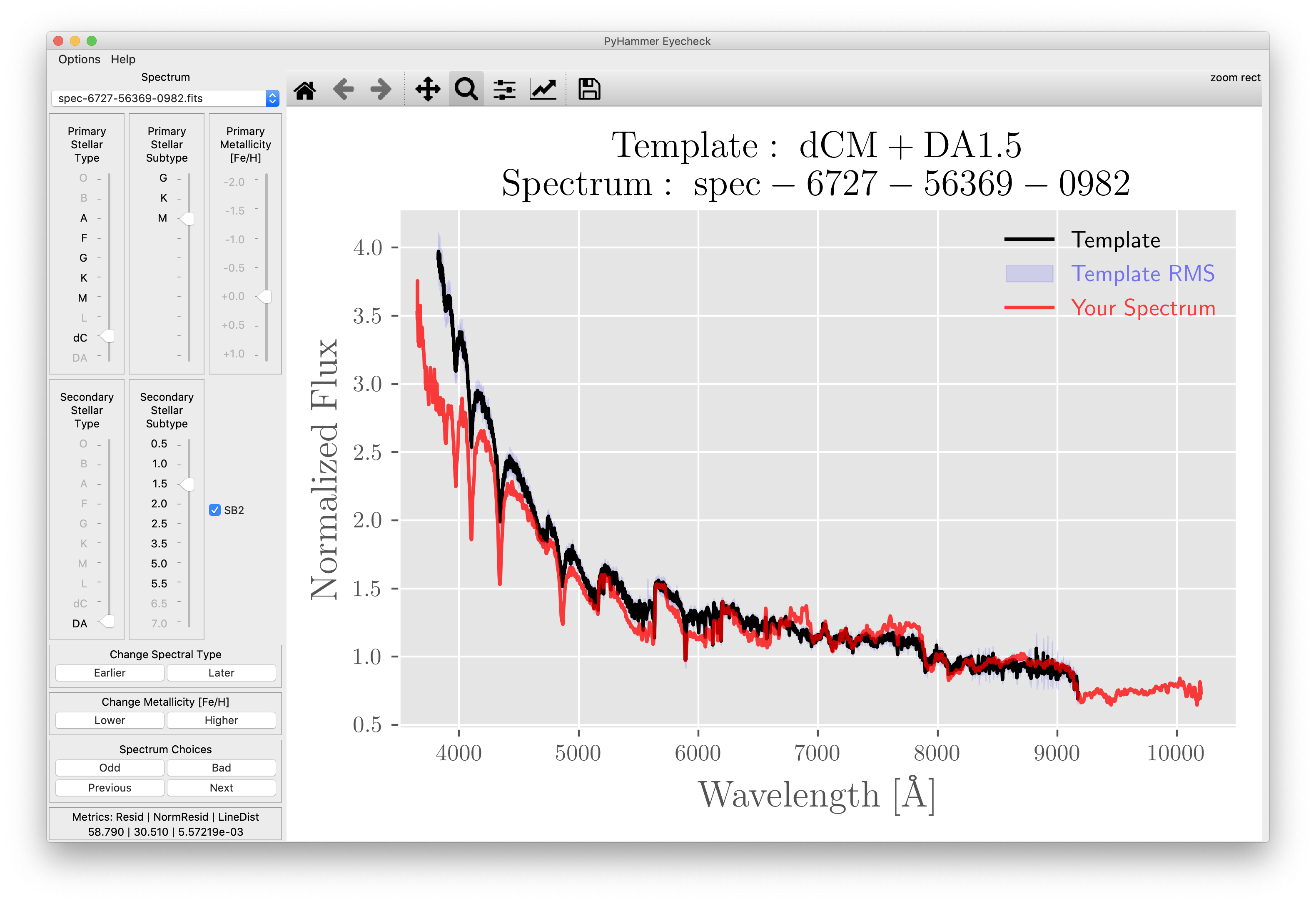}
\caption{The GUI for PyHammer v2.0. This GUI is functionally the same as in the original PyHammer. The main new feature is that of the addition of SB2 options.  Users now have the option to toggle on and off SB2 templates with a check button. This button enables and disables the SB2 sliders which allow the user to choose specific combinations of available SB2 binaries. This allows users to compare the fits of SB2 and single stars as well as to change the SB2 secondary types and sub-types to check for best fits.}
\label{fig:gui}
\end{figure*}

 \section{Summary} \label{summary}
We have extended the PyHammer spectral typing software to include new carbon and DA white dwarf single star templates. These new templates were created in a similar method as the original PyHammer stellar templates via the co-addition of SDSS optical spectra. These new templates cover a range of effective temperatures across both dC and DA WD classes, providing spectral typing abilities for unique and important stellar types.  

In addition, we have also created a new luminosity-normalized spectral library that consists of stars across the MK classification types. These luminosity templates are based on two libraries of accurately flux calibrated optical spectra and using the Gaia DR2 to convert to luminosity units of erg s$^{-1}$ \AA$^{-1}$. This luminosity library allowed us to create combinations of double-lined spectroscopic binary (SB2) templates which we have also added to this v2.0 of PyHammer.

Fast and accurate automatic spectral typing is important for individual observers but also for large scale all-sky surveys of today and the future. Surveys such as the SDSS-IV \citep{SDSS_4} and the upcoming SDSS-V \citep{SDSS_5} need accurate stellar spectral types in their reduction pipelines. These surveys often use stellar templates based on synthetic spectra and model atmospheres that require assumptions and simplifications. The stellar templates presented here allow for accurate spectral typing for situations in which accurate stellar models do not exist and would normally be left out of synthetic template libraries, such as dwarf carbon stars 
\citep{Green2013}. 

PyHammer is also easily extendable to any spectral class in the future. The requirements being only that there exist enough correctly typed spectra to create a template and for which there are measurable spectral line features characteristic of that type. Examples of future PyHammer extensions could be CVs, T Tauri stars, QSOs, or other classes of galaxies. 
It is also possible that PyHammer could be extended to other wavelengths, for example to encompass infrared (IR) spectra. This type of extension would only require additional spectral indices in the desired wavelength ranges and would be useful for IR spectral surveys such as APOGEE \citep{APOGEE}.

\facility{Sloan 2.5-meter}

\software{\texttt{astropy} \citep{astropy}, \texttt{matplotlib} \citep{matplotlib}, \texttt{numpy} \citep{numpy}, \texttt{PyHammer} \citep{PyHammer_1}, \texttt{scipy} \citep{scipy}}


\acknowledgments
{Funding for the Sloan Digital Sky Survey IV has been provided by the Alfred P. Sloan Foundation, the U.S. Department of Energy Office of Science, and the Participating Institutions. SDSS-IV acknowledges support and resources from the Center for High-Performance Computing at the University of Utah. The SDSS web site is www.sdss.org.

SDSS-IV is managed by the Astrophysical Research Consortium for the Participating Institutions of the SDSS Collaboration including the Brazilian Participation Group, the Carnegie Institution for Science, Carnegie Mellon University, the Chilean Participation Group, the French Participation Group, Center for Astrophysics $\vert$ Harvard \& Smithsonian, Instituto de Astrof\'isica de Canarias, The Johns Hopkins University, Kavli Institute for the Physics and Mathematics of the Universe (IPMU) / University of Tokyo, the Korean Participation Group, Lawrence Berkeley National Laboratory, Leibniz Institut f\"ur Astrophysik Potsdam (AIP),  Max-Planck-Institut f\"ur Astronomie (MPIA Heidelberg), Max-Planck-Institut f\"ur Astrophysik (MPA Garching), Max-Planck-Institut f\"ur Extraterrestrische Physik (MPE), National Astronomical Observatories of China, New Mexico State University, New York University, University of Notre Dame, Observat\'ario Nacional / MCTI, The Ohio State University, Pennsylvania State University, Shanghai Astronomical Observatory, United Kingdom Participation Group, Universidad Nacional Aut\'onoma de M\'exico, University of Arizona, University of Colorado Boulder, University of Oxford, University of Portsmouth, University of Utah, University of Virginia, University of Washington, University of Wisconsin, Vanderbilt University, and Yale University.}

\bibliographystyle{aasjournal}
\bibliography{references}{}


\end{document}